\documentclass[12pt]{article}
\usepackage{amsmath,amssymb,bm,graphicx}
\usepackage{graphicx}
\usepackage[dvips]{color} 

\begin{document}

\makeatletter
\@addtoreset{equation}{section}
\def\theequation{\thesection.\arabic{equation}}
\makeatother

\vskip 0.5 truecm
\vskip 0.5 truecm

\begin{center}
{\Large{\bf A hidden-variables version of Gisin's theorem
}}
\end{center}
\vskip .5 truecm
\centerline{\bf  Kazuo Fujikawa$^1$ and Koichiro Umetsu$^{2}$}
\vskip .4 truecm
\centerline {\it $^1$ Mathematical Physics Laboratory,}
\centerline {\it RIKEN Nishina Center, Wako 351-0198, Japan}
\vspace{0.3cm}
\centerline {\it $^2$ 
Laboratory of Physics, College of Science and Technology, }
\centerline {\it  Nihon University,  Funabashi, Chiba 274-8501, Japan
}
\vskip 0.5 truecm

\begin{abstract}
It is generally assumed that {\em local realism} represented by a noncontextual and local hidden-variables model  in $d=4$  such as the one used by Bell always gives rise to  CHSH inequality $|\langle B\rangle|\leq 2$. On the other hand, the contraposition of Gisin's theorem states that the inequality $|\langle B\rangle|\leq 2$ for arbitrary parameters implies (pure) separable quantum  states.  The fact that local realism can describe only pure separable quantum states is  naturally established in hidden-variables models, and it is quantified by $G({\bf a},{\bf b})= 4[\langle \psi|P({\bf a})\otimes P({\bf b})|\psi\rangle-\langle \psi|P({\bf a})\otimes{\bf 1}|\psi\rangle\langle \psi|{\bf 1}\otimes P({\bf b})|\psi\rangle]=0$ for any two projection operators $P({\bf a})$ and $P({\bf b})$.  The test of local realism by the deviation of $G({\bf a},{\bf b})$ from 
$G({\bf a},{\bf b})=0$ is shown to be very efficient using the past experimental setup of Aspect and his collaborators in 1981.
\end{abstract}


\section{Introduction}
Entanglement in quantum mechanics intuitively implies some form of non-locality, and the locality issue  is usually discussed in terms of the notion of local realism. Although a precise mathematical definition of local realism appears to be absent, it is customary to use  noncontextual and  local hidden-variables models as a concrete realization of local realism.
In the present paper we follow this tradition,  and we use  the hidden-variables model that belongs to the same class of models as in the original paper of Bell~\cite{bell} which is noncontextual and local (i.e., a noncontextual model applied to only far apart parties) as classified by Mermin\cite{mermin}. 

The purpose of the present paper is to propose a very simple test of local realism in addition to the well-known CHSH inequality~\cite{chsh}. This criterion is based on the fact that  {\em non-contextual and local hidden-variables models in $d=2\times2=4$, namely, local realism can describe only separable quantum mechanical states}. This criterion is quantified by,
\begin{eqnarray}
G({\bf a},{\bf b})\equiv 4[\langle \psi|P({\bf a})\otimes P({\bf b})|\psi\rangle-\langle \psi|P({\bf a})\otimes{\bf 1}|\psi\rangle\langle \psi|{\bf 1}\otimes P({\bf b})|\psi\rangle]=0, 
\end{eqnarray}
for arbitrary two projection operators $P({\bf a})$ and $ P({\bf b})$. A decisive test of local realism is then given by the deviation of $G({\bf a},{\bf b})$ from $G({\bf a},{\bf b})=0$. 
We suspect that this fact itself has been known,  but,  to our knowledge,  no systematic use of this fact as a practical test of local realism has been attempted in the past; to substantiate this statement, we later discuss the early experiment by Aspect and his collaborators in 1981~\cite{aspect1}.
 
We are going to first derive the above fact directly by an analysis of
local hidden-variables models. Here the linearity of the probability measure, which is fundamental to Born's probability interpretation of  quantum mechanics, is crucial. Then we argue that the same result is inferred from a combination of (the contraposition of) Gisin's theorem in quantum mechanics~\cite{werner, gisin} with the well-known Bell's theorem in hidden-variables models; this implies that the linearity of the probability measure is implicitly assumed in the derivation of Bell-CHSH inequality\cite{chsh}. 
We here recall that the basic operational rule of the hidden-variables model is to translate a quantum mechanical
statement into the language of the hidden-variables model and then after the manipulations allowed in the hidden-variables model, translate the final result back into a statement of quantum mechanics which may be tested by experiments.  Our relation $G({\bf a},{\bf b})=0$ in (1.1) and the well-known CHSH inequality $|\langle B\rangle|\leq 2$~\cite{chsh}, both of which are stated in quantum mechanical language, are the consequences of local realism in this sense.

\section{Hidden-variables model}
We start with the assumption that the correlation for two projection operators $A_{1}$ and $B_{2}$ belonging to systems  
$1$ and $2$, respectively, is expressed in $d=4$ as
\begin{eqnarray}
\langle \psi|A_{1}\otimes B_{2}|\psi\rangle=\int \rho(\lambda_{1},\lambda_{2})d\lambda_{1}d\lambda_{2}A_{1}(\psi,\lambda_{1})B_{2}(\psi,\lambda_{2}),
\end{eqnarray}
where $A_{1}(\psi,\lambda_{1})$ and $B_{2}(\psi,\lambda_{2})$ assume the eigenvalues of projectors, namely, $1$ or $0$.
This specific hidden-variables representation was mentioned by Bell in his original paper~\cite{bell}; he argued that this representation is included in the more conventional one 
\begin{eqnarray}
\langle \psi|A_{1}\otimes B_{2}|\psi\rangle=\int \rho(\lambda)d\lambda A_{1}(\psi,\lambda)B_{2}(\psi,\lambda),
\end{eqnarray}
if one writes $d\lambda=d\lambda_{1}d\lambda_{2}$ and recalls the fact that the $\lambda$ dependence of the other factors in (2.2) is arbitrary. The converse is also true since one recovers the more conventional representation (2.2) from (2.1) by setting
\begin{eqnarray}
\rho(\lambda_{1},\lambda_{2})=\rho(\lambda_{1})\delta(\lambda_{1}-\lambda_{2}).
\end{eqnarray}
The representation (2.1) is convenient for our purpose. The representation (2.1) is noncontexual in the sense that $\rho(\lambda_{1},\lambda_{2})$ is independent of the choice of variables $A_{1}$ and $B_{2}$ and the state $|\psi\rangle$. Also
$A_{1}(\psi,\lambda_{1})$ is independent of the partner $B_{2}(\psi,\lambda_{2})$, and vice versa. As for the $\psi$ dependence of $A_{1}(\psi,\lambda_{1})$ and $B_{2}(\psi,\lambda_{2})$, we follow the concrete examples in 
$d=2$~\cite{bell2, beltrametti}; on the other hand, in the original model of Bell in $d=4$~\cite{bell} the $\psi$ dependence is included in hidden-variables. It is confirmed that this difference  does not modify our analysis and conclusion.

In defining Clauser-Horne-Shimony-Holt (CHSH) inequality for two spin systems~\cite{chsh}, it is natural to consider 
the quantum mechanical operator $B$ introduced by Cirel'son~\cite{cirel'son}
\begin{eqnarray}
B={\bf a}\cdot {\bf \sigma}\otimes ({\bf b}+{\bf b}^{\prime})\cdot {\bf \sigma} +{\bf a}^{\prime}\cdot{\bf \sigma}\otimes ({\bf b}-{\bf b}^{\prime})\cdot{\bf \sigma},
\end{eqnarray}
where ${\bf \sigma}$ stands for the Pauli matrix and  ${\bf a},\ {\bf a}^{\prime},\ {\bf b},\ {\bf b}^{\prime}$ are 3-dimensional unit vectors. This operator is bounded by 
\begin{eqnarray}
||B||\leq |{\bf b}+{\bf b}^{\prime}|+|{\bf b}-{\bf b}^{\prime}|\leq 2\sqrt{2}.
\end{eqnarray}
One  has the quantum mechanically equivalent  relations
\begin{eqnarray}
\langle B\rangle&=&\langle {\bf a}\cdot {\bf \sigma}\otimes ({\bf b}+{\bf b}^{\prime})\cdot {\bf \sigma}\rangle +\langle {\bf a}^{\prime}\cdot{\bf \sigma}\otimes ({\bf b}-{\bf b}^{\prime})\cdot{\bf \sigma}\rangle\nonumber\\
&=&\langle {\bf a}\cdot {\bf \sigma}\otimes {\bf b}\cdot {\bf \sigma}\rangle +
\langle{\bf a}\cdot {\bf \sigma}\otimes {\bf b}^{\prime}\cdot {\bf \sigma}\rangle\nonumber\\
&& +\langle {\bf a}^{\prime}\cdot{\bf \sigma}\otimes {\bf b}\cdot{\bf \sigma}\rangle-\langle {\bf a}^{\prime}\cdot{\bf \sigma}\otimes {\bf b}^{\prime}\cdot{\bf \sigma}\rangle. 
\end{eqnarray}
The hidden-variables representation is then defined by 
\begin{eqnarray}
\langle \psi|{\bf a}\cdot {\bf \sigma}\otimes {\bf b}\cdot {\bf \sigma}|\psi\rangle=\int \rho(\lambda_{1},\lambda_{2})d\lambda_{1}d\lambda_{2}a(\psi,\lambda_{1})b(\psi,\lambda_{2})
\end{eqnarray} 
with dichotomic variables $a(\psi,\lambda_{1})$ and $b(\psi,\lambda_{2})$ which assume the eigenvalues of ${\bf a}\cdot {\bf \sigma}$ and ${\bf b}\cdot {\bf \sigma}$, namely, 
$\pm 1$. This is based on the spectral decomposition such as ${\bf a}\cdot {\bf \sigma}=c_{1}P_{1}+c_{2}P_{2}$ with projectors $P_{1}+P_{2}=1$ and $P_{1}P_{2}=0$. On the other hand, the spectral decomposition implies
\begin{eqnarray}
\langle \psi| {\bf a}\cdot {\bf \sigma}\otimes ({\bf b}+{\bf b}^{\prime})\cdot {\bf \sigma}|\psi\rangle&=&|{\bf b}+{\bf b}^{\prime}|\langle \psi| {\bf a}\cdot {\bf \sigma}\otimes \tilde{{\bf b}}\cdot {\bf \sigma}|\psi\rangle\\
&=&|{\bf b}+{\bf b}^{\prime}|\int \rho(\lambda_{1},\lambda_{2})d\lambda_{1}d\lambda_{2}a(\psi,\lambda_{1})\tilde{b}(\psi,\lambda_{2})\nonumber
\end{eqnarray} 
with a unit vector $\tilde{{\bf b}}\equiv ({\bf b}+{\bf b}^{\prime})/|{\bf b}+{\bf b}^{\prime}|$
for non-collinear ${\bf b}$ and ${\bf b}^{\prime}$; note that this operation is consistent with assumed {\em locality}.

If one moves to the hidden-variables representation from the last expression in (2.6), one obtains the standard CHSH relation~\cite{chsh}, $|\langle B\rangle|\leq 2$,
\begin{eqnarray}
&&|\langle \psi| {\bf a}\cdot {\bf \sigma}\otimes {\bf b}\cdot {\bf \sigma}|\psi\rangle +
\langle \psi|{\bf a}\cdot {\bf \sigma}\otimes {\bf b}^{\prime}\cdot {\bf \sigma}|\psi\rangle\nonumber\\
&& +\langle \psi| {\bf a}^{\prime}\cdot{\bf \sigma}\otimes {\bf b}\cdot{\bf \sigma}|\psi\rangle-\langle \psi| {\bf a}^{\prime}\cdot{\bf \sigma}\otimes {\bf b}^{\prime}\cdot{\bf \sigma}|\psi\rangle|\leq 2,
\end{eqnarray}
by noting,
\begin{eqnarray}
&&a(\psi,\lambda_{1})b(\psi,\lambda_{2})+a(\psi,\lambda_{1})b^{\prime}(\psi,\lambda_{2})\nonumber\\
&&+a^{\prime}(\psi,\lambda_{1})b(\psi,\lambda_{2})
-a^{\prime}(\psi,\lambda_{1})b^{\prime}(\psi,\lambda_{2})=\pm 2,
\end{eqnarray}
 for dichotomic variables such as $a(\psi,\lambda_{1})=\pm 1$, while if one moves to the hidden-variables representation from the first expression in (2.6)  one {\em cannot prove} $|\langle B\rangle|\leq 2$ in general~\cite{fujikawa} and thus only the general quantum mechanical bound $|\langle B\rangle|\leq 2\sqrt{2}$ as in (2.5). To achieve the conventional CHSH inequality $|\langle B\rangle|\leq 2$ uniquely, one needs to satisfy the linearity of the probability measure 
\begin{eqnarray}
\langle {\bf a}\cdot {\bf \sigma}\otimes ({\bf b}+{\bf b}^{\prime})\cdot {\bf \sigma}\rangle=\langle {\bf a}\cdot {\bf \sigma}\otimes {\bf b}\cdot {\bf \sigma}\rangle +
\langle{\bf a}\cdot {\bf \sigma}\otimes {\bf b}^{\prime}\cdot {\bf \sigma}\rangle
\end{eqnarray}
for any ${\bf a}\cdot {\bf \sigma}$ including ${\bf a}\cdot {\bf \sigma}$ replaced by the unit operator ${\bf 1}$ and for any non-collinear ${\bf b}$ and ${\bf b}^{\prime}$  in the hidden-variables representation; similarly, the relations with ${\bf a}$ and ${\bf b}$ interchanged. Quite apart from CHSH inequality, this linearity of the probability measure (2.11), which is fundamental to the formulation of Born probability interpretation and leads to the density matrix representation~\cite{neumann}, is a natural requirement; the known concrete hidden-variables models in $d=2$~\cite{bell2, kochen} satisfy the  relation,
\begin{eqnarray}
\langle {\bf 1} \otimes ({\bf b}+{\bf b}^{\prime})\cdot {\bf \sigma}\rangle=\langle {\bf 1}\otimes  {\bf b}\cdot {\bf \sigma}\rangle +
\langle {\bf 1}\otimes {\bf b}^{\prime}\cdot {\bf \sigma}\rangle
\end{eqnarray}
when this is regarded as a relation in $d=2$ since these models reproduce quantum mechanics,
and thus the hidden-variables model in $d=4$ should naturally satisfy the relation (2.11) if it should reproduce some essential aspects of quantum mechanics. 

This requirement (2.11) imposes a strong condition on hidden-variables models:
\begin{eqnarray}
&&|{\bf b}+{\bf b}^{\prime}|\int \rho(\lambda_{2})d\lambda_{2}\tilde{b}(\psi,\lambda_{2})\nonumber\\
&&=\int \rho(\lambda_{2})d\lambda_{2}b(\psi,\lambda_{2})+\int \rho(\lambda_{2})d\lambda_{2}b^{\prime}(\psi,\lambda_{2})
\end{eqnarray}
and
\begin{eqnarray}
&&|{\bf b}+{\bf b}^{\prime}|\int \rho(a,\psi;\lambda_{2})d\lambda_{2}\tilde{b}(\psi,\lambda_{2})\nonumber\\
&&=\int \rho(a,\psi;\lambda_{2})d\lambda_{2}b(\psi,\lambda_{2})+\int \rho(a,\psi;\lambda_{2})d\lambda_{2}b^{\prime}(\psi,\lambda_{2})
\end{eqnarray}
where
\begin{eqnarray}
&&\rho(\lambda_{2})\equiv\int \rho(\lambda_{1},\lambda_{2})d\lambda_{1},\nonumber\\
&&\rho(a,\psi;\lambda_{2})\equiv\int \rho(\lambda_{1},\lambda_{2})\frac{1}{2}[1-a(\psi,\lambda_{1})]d\lambda_{1}.
\end{eqnarray}
The relation (2.13) shows that the weight factor $\rho(\lambda_{2})$ in (2.15)
defines a $d=2$ noncontextual hidden variables model for $b(\psi,\lambda_{2})$.
Note that the relation for {\em each} $\lambda_{2}$
\begin{eqnarray}
|{\bf b}+{\bf b}^{\prime}|\tilde{b}(\psi,\lambda_{2})
=b(\psi,\lambda_{2})+b^{\prime}(\psi,\lambda_{2})
\end{eqnarray}
does not hold for non-collinear ${\bf b}$ and ${\bf b}^{\prime}$ due to the well-known argument of von Neumann~\cite{neumann}.
Similarly, the relation (2.14), which is defined by a difference between (2.11) and (2.12), shows that the non-negative $\rho(a,\psi;\lambda_{2})$  in (2.15) defines a $d=2$ hidden variables model for the same presentation $b(\psi,\lambda_{2})$; namely, $a(\psi,\lambda_{1})$ parameterizes the functional form of the weight factor $\rho(a,\psi;\lambda_{2})\geq 0$. This shows that $d=2$ hidden-variables model defined by $b(\psi,\lambda_{2})$ allows a continuous number of weight factors. It may be natural to assume that the weight factor is uniquely defined for a given representation $b(\psi,\lambda_{2})$, which is the case of the models of Bell~\cite{bell2, beltrametti} and Kochen-Specker~\cite{kochen} in $d=2$, or at least one may ask no continuous degeneracy of the weight factor. In this case, one concludes that 
\begin{eqnarray}
\rho(\lambda_{1},\lambda_{2})=\rho_{1}(\lambda_{1})\rho_{2}(\lambda_{2})
\end{eqnarray}
where $\rho_{1}$ and $\rho_{2}$ stand for weight factors for  $d=2$ models, which satisfy the linearity of probability measure (2.12), such as given by Bell and Kochen-Specker. The factors $\rho(\lambda_{2})$ and $\rho(a,\psi;\lambda_{2})$ in (2.15) then become equivalent as a weight factor for $b(\psi,\lambda_{2})$. 
Any $d=4$ non-contextual and local hidden-variables model, when the linearity of probability measure (2.12) is asked, is thus represented by a factored product of $d=2$ models~\cite{fujikawa, fuji-oh-zhang}, 
\begin{eqnarray}
\langle \psi|{\bf a}\cdot {\bf \sigma}\otimes {\bf b}\cdot {\bf \sigma}|\psi\rangle=\int \rho_{1}(\lambda_{1})d\lambda_{1}a(\psi,\lambda_{1})\int \rho_{2}(\lambda_{2})d\lambda_{2}
b(\psi,\lambda_{2}).
\end{eqnarray}
This relation, when translated into the language of quantum mechanics, implies
\begin{eqnarray}
\langle \psi|{\bf a}\cdot {\bf \sigma}\otimes {\bf b}\cdot {\bf \sigma}|\psi\rangle=\langle \psi|{\bf a}\cdot {\bf \sigma}\otimes{\bf 1}|\psi\rangle\langle \psi|{\bf 1}\otimes{\bf b}\cdot {\bf \sigma}|\psi\rangle.
\end{eqnarray}
The relations (2.18) and (2.19) show that non-contextual and local hidden-variables models  in $d=4$ (local realism) can describe only the separable pure quantum states~\cite{fujikawa, fuji-oh-zhang}. This analysis goes through using projection operators~\cite{fuji-oh-zhang} and thus it is valid for a general $d=4$ system such as a two-photon system. We have thus established $G({\bf a},{\bf b})=0$ in (1.1) as a prediction of local realism.

The above conclusion (2.18) is compared  with (the contraposition of) ordinary Gisin's theorem which states that $|\langle \psi| B|\psi\rangle|\leq 2$ for arbitrary ${\bf a},\ {\bf a}^{\prime},\ {\bf b},\ {\bf b}^{\prime}$ implies separable states if one considers only the pure states in $d=4$~\cite{werner, gisin}; we emphasize that Gisin's theorem is a quantum mechanical statement and Gisin's theorem by itself has nothing to do with hidden-variables models. On the other hand, the well-known Bell's theorem states that the noncontextual and local hidden-variables model gives rise to 
$|\langle \psi| B|\psi\rangle|\leq 2$ for arbitrary ${\bf a},\ {\bf a}^{\prime},\ {\bf b},\ {\bf b}^{\prime}$. These two statements put together are consistent with our relation (2.18), and this agreement implies that the derivation of Bell-CHSH inequality implicitly assumes the linearity of the probability measure (2.11), on which the quantum mechanical Gisin's theorem also crucially depends.  We tentatively call (2.18) as {\em a hidden-variables version of Gisin's theorem}, since it was derived in hidden-variables models. Note that we here consider only the pure states in the sense of $d=4$. 

The relation (2.18), which is equivalent to (2.19), does not lead to any difficulty in connection with "contextuality" noted by Mermin~\cite{mermin}; for example, the valuations (which assign eigenvalues to operators) $v(\sigma^{1}_{x}\sigma^{2}_{y})$, $v(\sigma^{1}_{y}\sigma^{2}_{x})$, $v(\sigma^{1}_{z}\sigma^{2}_{z})$, where $\vec{\sigma}^{1}$ and $\vec{\sigma}^{2}$ are spin operators of the system 1 and 2 respectively, are regarded
as valuations of 3 commuting operators in~\cite{mermin}, but these correspond to the valuations of non-commuting operators 
$v_{1}(\sigma^{1}_{x})v_{2}(\sigma^{2}_{y})$,\ $v_{1}(\sigma^{1}_{y})v_{2}(\sigma^{2}_{x})$,\ $v_{1}(\sigma^{1}_{z})v_{2}(\sigma^{2}_{z})$,
in the context of (2.18) and thus cannot be used in the argument of contextuality.

\section{Tests of local realism}

We can categorize three different cases:\\
(i) Quantum mechanics. We can define
\begin{eqnarray}
\langle \psi| B|\psi\rangle_{QM}&=&\langle \psi|{\bf a}\cdot {\bf \sigma}\otimes {\bf b}\cdot {\bf \sigma}|\psi\rangle +
\langle \psi|{\bf a}\cdot {\bf \sigma}\otimes {\bf b^{\prime}}\cdot {\bf \sigma}|\psi\rangle\nonumber\\
&& +\langle \psi|{\bf a^{\prime}}\cdot {\bf \sigma}\otimes {\bf b}\cdot {\bf \sigma}|\psi\rangle-\langle \psi|{\bf a^{\prime}}\cdot {\bf \sigma}\otimes {\bf b^{\prime}}\cdot {\bf \sigma}|\psi\rangle
\end{eqnarray}
which assumes values $|\langle \psi| B|\psi\rangle_{QM}|\leq 2\sqrt{2}$.\\
(ii) Bell-CHSH inequality. We have a prediction of a noncontextual and local hidden-variables model (local realism)
\begin{eqnarray}
|\langle \psi| B|\psi\rangle_{CHSH}|&=&|\langle \psi|{\bf a}\cdot {\bf \sigma}\otimes {\bf b}\cdot {\bf \sigma}|\psi\rangle +
\langle \psi|{\bf a}\cdot {\bf \sigma}\otimes {\bf b^{\prime}}\cdot {\bf \sigma}|\psi\rangle\nonumber\\
&& +\langle \psi|{\bf a^{\prime}}\cdot {\bf \sigma}\otimes {\bf b}\cdot {\bf \sigma}|\psi\rangle-\langle \psi|{\bf a^{\prime}}\cdot {\bf \sigma}\otimes {\bf b^{\prime}}\cdot {\bf \sigma}|\psi\rangle|\nonumber\\
&\leq& 2,
\end{eqnarray}
and the emphasis here is that this is valid for {\em any state} $|\psi\rangle$. We can thus test if a given state is described by local realism.\\
(iii) A hidden-variables version of Gisin's theorem. A noncontextual and local hidden-variables model (local realism) describes only separable states and thus predicts
\begin{eqnarray}
\langle \psi|{\bf a}\cdot {\bf \sigma}\otimes {\bf b}\cdot {\bf \sigma}|\psi\rangle&=&\langle \psi|{\bf a}\cdot {\bf \sigma}\otimes {\bf 1}|\psi\rangle\langle \psi|{\bf 1}\otimes {\bf b}\cdot {\bf \sigma}|\psi\rangle,
\end{eqnarray}
and the emphasis here is that this is valid for {\em any choice} of ${\bf a}$ and ${\bf b}$, including
\begin{eqnarray}
\langle \psi|{\bf a}\cdot {\bf \sigma}\otimes {\bf b^{\prime}}\cdot {\bf \sigma}|\psi\rangle&=&\langle \psi|{\bf a}\cdot {\bf \sigma}\otimes {\bf 1}|\psi\rangle\langle \psi|{\bf 1}\otimes {\bf b^{\prime}}\cdot {\bf \sigma}|\psi\rangle, \nonumber\\
\langle \psi|{\bf a^{\prime}}\cdot {\bf \sigma}\otimes {\bf b}\cdot {\bf \sigma}|\psi\rangle&=&\langle \psi|{\bf a^{\prime}}\cdot {\bf \sigma}\otimes {\bf 1}|\psi\rangle\langle \psi|{\bf 1}\otimes {\bf b}\cdot {\bf \sigma}|\psi\rangle, \nonumber\\
\langle \psi|{\bf a^{\prime}}\cdot {\bf \sigma}\otimes {\bf b^{\prime}}\cdot {\bf \sigma}|\psi\rangle&=&\langle \psi|{\bf a^{\prime}}\cdot {\bf \sigma}\otimes {\bf 1}|\psi\rangle\langle \psi|{\bf 1}\otimes {\bf b^{\prime}}\cdot {\bf \sigma}|\psi\rangle.
\end{eqnarray}
By asking the relation (3.3) for any choice of ${\bf a}$ and ${\bf b}$, one can test if a given state is described by local realism.
 
We here recall the well-known fact in quantum mechanics that 
$-2\leq\langle \psi| B|\psi\rangle_{QM}\leq 2$ for any separable state $|\psi\rangle=|\psi_{1}\rangle|\psi_{2}\rangle$. This is because 
\begin{eqnarray}
-2&\leq& \langle \psi_{1}|{\bf a}\cdot {\bf \sigma}|\psi_{1}\rangle\langle \psi_{2}|{\bf b}\cdot {\bf \sigma}|\psi_{2}\rangle + \langle \psi_{1}|{\bf a}\cdot {\bf \sigma}|\psi_{1}\rangle\langle \psi_{2}|{\bf b^{\prime}}\cdot {\bf \sigma}|\psi_{2}\rangle\nonumber\\
&+& \langle \psi_{1}|{\bf a^{\prime}}\cdot {\bf \sigma}|\psi_{1}\rangle\langle \psi_{2}|{\bf b}\cdot {\bf \sigma}|\psi_{2}\rangle - \langle \psi_{1}|{\bf a^{\prime}}\cdot {\bf \sigma}|\psi_{1}\rangle\langle \psi_{2}|{\bf b^{\prime}}\cdot {\bf \sigma}|\psi_{2}\rangle\leq 2, 
\end{eqnarray}
since $\langle \psi| B|\psi\rangle_{QM}$ is linear in all the variables, for example, $\langle \psi_{1}|{\bf a}\cdot {\bf \sigma}|\psi_{1}\rangle$ and $\langle \psi_{1}|{\bf a^{\prime}}\cdot {\bf \sigma}|\psi_{1}\rangle$ with $-1\leq \langle \psi_{1}|{\bf a}\cdot {\bf \sigma}|\psi_{1}\rangle\leq +1$ and $-1\leq\langle \psi_{1}|{\bf a^{\prime}}\cdot {\bf \sigma}|\psi_{1}\rangle\leq +1$. The linear function $\langle \psi| B|\psi\rangle_{QM}$ becomes maximum or minimum at the boundary of the domain, and one can check four corners $(\langle \psi_{1}|{\bf a}\cdot {\bf \sigma}|\psi_{1}\rangle, \langle \psi_{1}|{\bf a^{\prime}}\cdot {\bf \sigma}|\psi_{1}\rangle)=(1,1),\ (1,-1),\ (-1,1),\\ 
\ (-1,-1)$ and find $-2\leq\langle \psi| B|\psi\rangle_{QM}\leq 2$. The separable state thus implies $|\langle \psi| B|\psi\rangle_{QM}|\leq 2$, and 
its contraposition states that $2<|\langle \psi| B|\psi\rangle_{QM}|\leq 2\sqrt{2}$ implies inseparable (entangled) states. On the other hand, Gisin's theorem states that $|\langle \psi| B|\psi\rangle_{QM}|\leq 2$ for all the possible ${\bf a},\ {\bf a}^{\prime},\ {\bf b},\ {\bf b}^{\prime}$ implies a separable state. All of these statements are quantum mechanical ones.

The noncontextual and local hidden-variables model (local realism) implies\\ $|\langle \psi| B|\psi\rangle_{CHSH}|\leq 2$ for all the states. Our criterion $G({\bf a},{\bf b})=0$ of local realism asks that all the quantum states be separable. We can thus test local realism.

Due to  (3.5), (iii) automatically implies (ii), while 
(ii) for  any ${\bf a},\ {\bf a}^{\prime},\ {\bf b},\ {\bf b}^{\prime}$ implies 
(iii) due to Gisin's theorem. In the analysis of actual experiments, however, we have only very limited combinations of parameters ${\bf a},\ {\bf a}^{\prime},\ {\bf b},\ {\bf b}^{\prime}$, thus 
the cases (ii) and (iii) generally give different constraints.
In the case (ii), one can judge if local realism is consistent with  inseparable  states by looking at the inequality, while  in the case (iii) one can judge if local realism is consistent with  inseparable  states by looking at the equality. 
Practically, a test of the single CHSH inequality (3.2) corresponds to the test of our criterion with 4 relations in (3.3) and (3.4).  Our criterion  thus tests the more detailed predictions for the same experimental setup. If one uses only a part of (3.3) and (3.4), our criterion
can in principle test local realism on the basis of a less number of experimental setups than required for the test of CHSH inequality.

The difference between Gisin's theorem and our hidden-variables version is also stated in the following way: In quantum mechanics, the pure states in $d=4$ are classified into two categories, namely,  for a given $B$, those states which  satisfy the condition $|\langle B\rangle|\leq 2$ and the rest which do not satisfy the condition, and those states which do not satisfy the condition are inseparable. Next we vary $B$, and we apply the same criterion to those states which  satisfy $|\langle B\rangle|\leq 2$
for the original $B$. The final result of this procedure is that  only those states which  satisfy $|\langle B\rangle|\leq 2$ for arbitrary choice of parameters in $B$ are separable. In comparison, any noncontextual and local hidden-variables model in $d=4$ describes only the pure separable states.
When the analytical expression of a state is given, the condition (3.3) thus  simply states that the state is literally separable, namely, any state
\begin{eqnarray}
\psi=\frac{1}{\sqrt{|\alpha|^{2}+|\beta|^{2}}}[\alpha|+\rangle_{1}|-\rangle_{2}-\beta|-\rangle_{1}|+\rangle_{2}],
\end{eqnarray}
where $|\pm\rangle_{1}$ and $|\pm\rangle_{2}$ stand for the eigenstates of $\sigma^{1}_{z}$ and $\sigma^{2}_{z}$, respectively, with $\alpha\beta\neq0$ cannot be compatible with local realism.

\section{Numerical illustration}

\vspace{0.3cm}
\begin{figure}[h]
  \begin{center}
    \includegraphics[width=80mm]{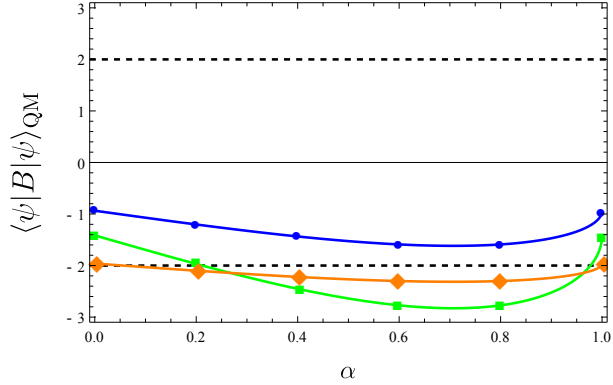}
  \end{center}
    \vspace{-0.3cm}
  \caption{{\bf Inseparable state and CHSH inequality}: We use the inseparable state (3.6) where both $\alpha$ and $\beta$ are real and positive with $\alpha^{2}+\beta^{2}=1$. Let ${\bf a}=(\sin\theta, 0, \cos \theta)$, ${\bf b}=(\sin\phi, 0, \cos \phi)$ and similarly for ${\bf a}'$ and ${\bf b}'$ by choosing y-axis in the direction of two separated parties. We performed numerical tests for three cases, namely, A: $(\theta,\phi,\theta',\phi')=(\frac{\pi}{3},\frac{\pi}{8},\frac{\pi}{4},\frac{\pi}{6})$, B: $(\frac{\pi}{4},\frac{\pi}{2},\frac{3\pi}{4},0)$, C: $(\frac{\pi}{6},\frac{3\pi}{4},\pi,0)$.
The lines with filled circle, square and diamond respectively indicate $\langle \psi | B|\psi \rangle_{{\rm QM}}$ in the case of A, B and C. 
The dashed lines stand for CHSH inequality (3.2), $-2\leq \langle \psi | B | \psi \rangle_{{\rm CHSH}} \leq 2$. }\label{fig01}
\end{figure}

\vspace{0.3cm}
\begin{figure}[h]
\begin{minipage}[c]{7.5cm}
    \includegraphics[width=70mm]{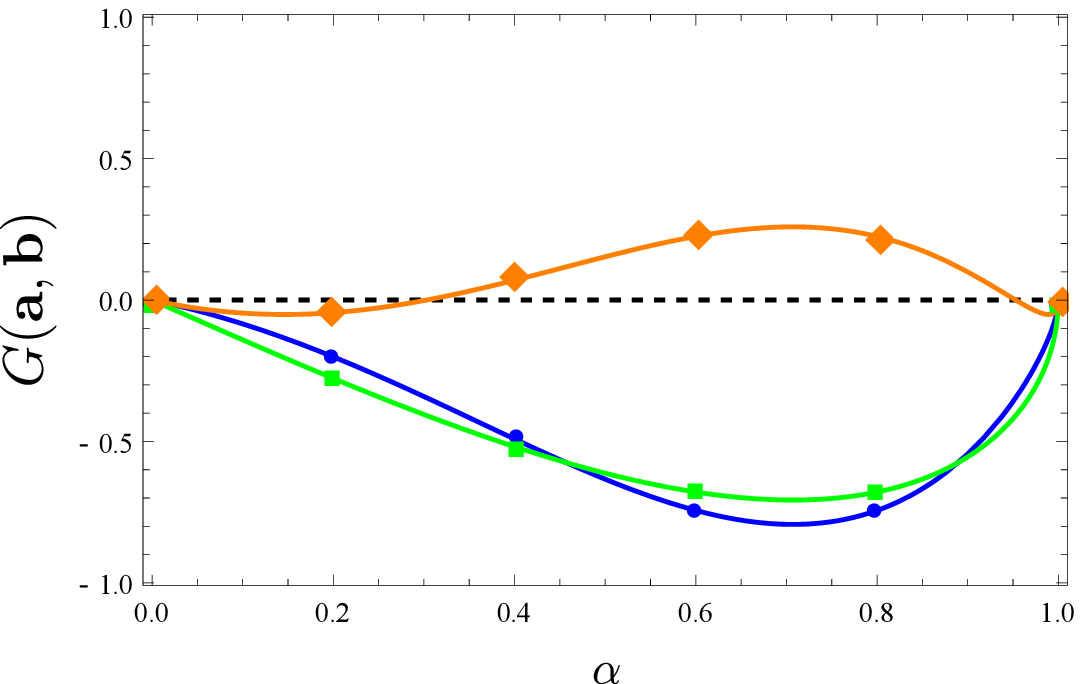}
\end{minipage}
\begin{minipage}[c]{7.5cm}
    \includegraphics[width=70mm]{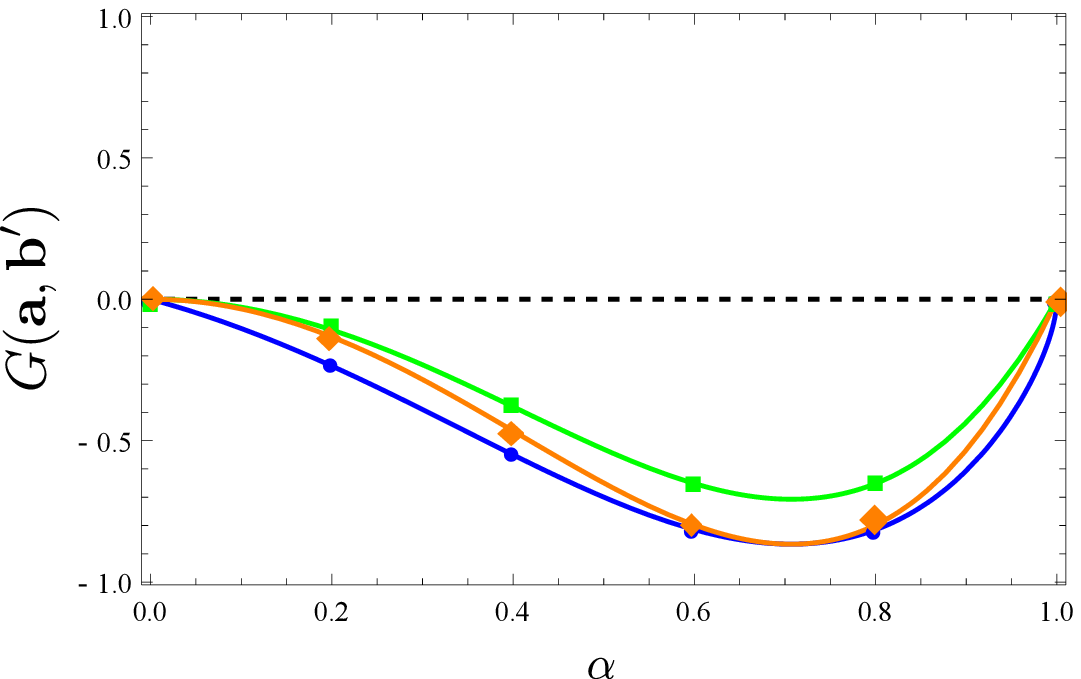}
\end{minipage}
    \vspace{-0.3cm}
   \caption{{\bf Inseparable state and a hidden-variable version of Gisin's theorem}: We use the same set of parameters as in Fig. 1. The lines with filled circle, square and diamond  indicate $G({\bf a},{\bf b})$ and $G({\bf a},{\bf b^{\prime}})$ defined in (4.1) for the case of A, B and C in Fig. 1, respectively. The dashed lines stand for the prediction of local realism. }
  \label{fig02}
\end{figure}

\vspace{0.4cm}
\begin{figure}[h]
  \begin{center}
    \includegraphics[width=80mm]{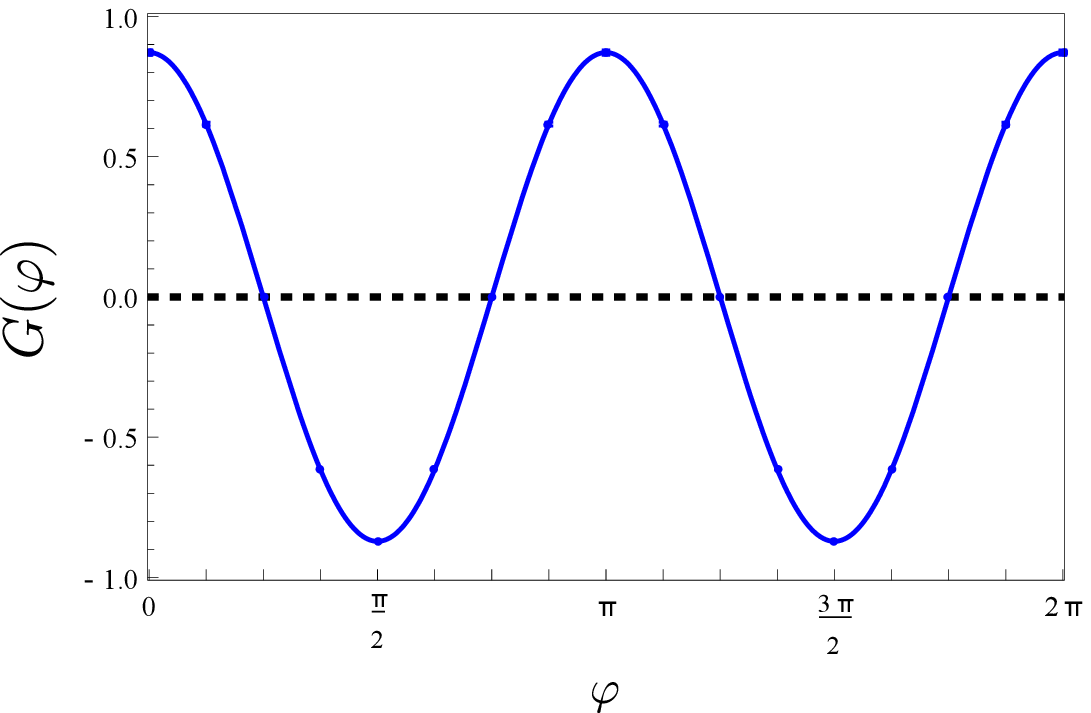}
  \end{center}
    \vspace{-0.3cm}
  \caption{{\bf A hidden-variables version of Gisin's theorem and Aspect's experiment}: The solid line represents the quantum mechanical prediction of $G(\varphi)$ defined in (4.2) corresponding to Aspect's experimental setup in 1981~\cite{aspect1}. The dashed line represents the prediction of local realism. }  \label{fig03}
\end{figure}

We first show $\langle \psi| B|\psi\rangle_{QM}$ in Fig. 1 for various experimental configurations. This figure shows that an inseparable pure state can satisfy CHSH inequality $|\langle \psi| B|\psi\rangle_{CHSH}|\leq 2$ for a certain parameter domain of the  inseparable  state. For such parameters, CHSH inequality cannot decide if the given state is  inseparable  or not. The choice of parameters in the case B is a well-known example that CHSH inequality is maximally violated, but
 one still sees that there exists a parameter domain, as is indicated by the line with square in Fig.1, for which CHSH inequality holds. Note that this is consistent with ordinary Gisin's theorem.

Similar analysis is performed for a hidden-variables version of Gisin's theorem in Fig. 2. To quantify the prediction of local realism (2.18), we plot
\begin{eqnarray}
G({\bf a},{\bf b})\equiv \langle \psi|{\bf a}\cdot {\bf \sigma}\otimes {\bf b}\cdot {\bf \sigma}|\psi\rangle-\langle \psi|{\bf a}\cdot {\bf \sigma}\otimes{\bf 1}|\psi\rangle\langle \psi|{\bf 1}\otimes{\bf b}\cdot {\bf \sigma}|\psi\rangle, 
\end{eqnarray}
for the same set of parameters as in Fig.1. This $G({\bf a},{\bf b})$ is equivalent to (1.1). 
We plot $G({\bf a},{\bf b})$ for (3.3) and $G({\bf a},{\bf b^{\prime}})$ for the first relation of (3.4), respectively. One sees that our criterion in (1.1) can differentiate local realism from quantum mechanics for all the given parameter ranges using a less number of experimental configurations. This illustrates that our criterion is more effective than CHSH inequality in negating local realism.

The experimental setups close to the test of our criterion in (1.1) have in fact been used in the past by Freedman and Clauser in 1972~\cite{freedman} and Aspect, Grangier and Roger in 1981~\cite{aspect1}. Those experiments are based on the measurement of the transverse linear polarization of the photon, while our analysis so far is based on two-spin systems. To rewrite our relations so that they are used for the photon measurement, we define the projection operator such as $P({\bf a})=(1+{\bf a}\cdot {\bf \sigma})/2$ or ${\bf a}\cdot {\bf \sigma}=2P({\bf a})-1$. The projector $P({\bf a})$ in the transverse direction is then identified with the photon linear polarizer in the direction ${\bf a}$. We then obtain the expression $G({\bf a},{\bf b})$ defined in (1.1).
In terms of measured quantities in~\cite{aspect1} , $G(\varphi)=G({\bf a},{\bf b})$ is written as  
\begin{eqnarray}
G(\varphi)=4[\frac{R(\varphi)}{R_0}-\frac{R_1R_2}{R_0^2}]=(0.971-0.029)(0.968-0.028)0.984 \cos2\varphi
\end{eqnarray}
where $\varphi$ stands for the angle between ${\bf a}$ and ${\bf b}$. (Note that for the maximally entangled state in quantum mechanics, we have $G({\bf a},{\bf b})=-\cos\varphi$ for the spin, while we have
$G({\bf a},{\bf b})=\cos2\varphi$
for the photon.)
The quantities $R(\varphi),\ R_1, \ R_2$ and $R_0$ are defined in eq.(2) of~\cite{aspect1}, and the numerical factors which appear in front of $\cos2\varphi$ are also given in~\cite{aspect1}. For the ideal measurement, the coefficient of $\cos2\varphi$ in (4.2) is unity. See also Refs.~\cite{chsh, freedman}. We show this quantum mechanical prediction in Fig.3 together with the prediction of local realism. Fig.3 shows that our criterion in (1.1) and (4.2) is very effective to test the deviation of local realism from quantum mechanics.
It should be emphasized that the authors in~\cite{aspect1} have not discussed  figure 4 in their paper, which corresponds to Fig.3 in the present paper, as a decisive test of local realism; instead they discussed only CHSH inequality as a test of local realism. This may show that our criterion (1.1) is not universally recognized as a decisive prediction of local realism.

\section{Conclusion}
We have illustrated the practical advantage of our criterion in (1.1) over CHSH inequality using the past experimental setups. It is rather surprising that, to our knowledge,  no systematic use of the extremely simple quantities (1.1) and (4.1) as a decisive test of local realism has been attempted in the past. Recently, more and more sophisticated tests
of Bell-CHSH inequalities have been performed~\cite{aspect2}-\cite{giustina}. It may be interesting to look at those 
experiments from the point of view of our criterion (1.1); since our criterion is mathematically much simpler, it may avoid some of the technical complications involved in the test of Bell-CHSH inequalities.

\section*{Acknowledgments}
One of the authors (K.F.) thanks C.H. Oh and Sixia Yu for helpful discussions at the Center for Quantum Technologies, National University of Singapore. 
This work is supported in part by JSPS KAKENHI (Grant No. 25400415).

\end{document}